\pdfoutput=1

\documentclass[conference]{IEEEtran}
\usepackage{array}
\usepackage{hhline, multirow}

\usepackage{cite}

\ifCLASSINFOpdf
  \usepackage[pdftex]{graphicx}
\else
  \usepackage[dvips]{graphicx}

\fi
\usepackage{amsmath}
\usepackage{amssymb}

\usepackage{algorithmic}
\ifCLASSOPTIONcompsoc
 \usepackage[caption=false,font=normalsize,labelfont=sf,textfont=sf]{subfig}
\else
 \usepackage[caption=false,font=footnotesize]{subfig}
\fi
\usepackage{url}

\begin{document}
\newcommand{\rom}[1]{\uppercase\expandafter{\romannumeral #1\relax}}
%
\title{Utilizing Probase in Open Directory Project-based Text Classification}

\author{\IEEEauthorblockN{So-Young Jun, Dinara Aliyeva, Ji-Min Lee, SangKeun Lee}
\IEEEauthorblockA{Korea University, Seoul, Republic of Korea\\
{\{syzzang, dinara\_aliyeva, jm94318, yalphy\}}@korea.ac.kr}}

\maketitle

\begin{abstract}
Open Directory Project (ODP) has been successfully utilized in text classification due to its representation ability of various categories. However, ODP includes a limited number of entities, which play an important role in classification tasks. In this paper, we enrich the semantics of ODP categories with Probase entities. To effectively incorporate Probase entities in ODP categories, we first represent each ODP category and Probase entity in terms of concepts. Next, we measure the semantic relevance between an ODP category and
a Probase entity based on the concept vector. Finally, we use Probase entity to enrich the semantics of the ODP categories. Our experimental results show that the proposed methodology exhibits a significant improvement over state-of-the-art techniques in the ODP-based text classification.
\end{abstract}


%
\IEEEpeerreviewmaketitle

\section{Introduction}

 
Text classification is the task of automatically assigning text to one or more categories. It plays an important role in many applications, such as contextual advertising \cite{lee2013semantic,ryu2017utilizing}, web search personalization \cite{ODP_personalized}, and personalized curation system \cite{ODP_Mobile1}. To capture various topics in arbitrary texts, text classification requires a sufficiently large taxonomy of topical categories \cite{shin2017utilizing} and a large amount of training data for each category.

In large-scale text classification, several studies \cite{lee2013semantic,ryu2017utilizing,shin2017utilizing} have utilized Open Directory Project (ODP)\footnote{\url{https://curlie.org}} due to its representation ability of various categories. The ODP contains approximately one million categories and millions of web pages, pre-classified into ODP categories. It is a large-scale web directory built by human editors. However, manually constructed knowledge bases exhibit a limited semantic information \cite{song2015automatic}. Especially the ODP suffers from the scarcity of entities, which may degrade the ODP-based text classification. 

An entity is a distinctly identifiable ``thing'', e.g., a specific person, event, or place, and is important to understand text \cite{Entity_query1, entityInQuery}. For example, given text ``I searched for Galaxy Nexus spec'', ``Galaxy Nexus" is the most important evidence that this text is relevant to cellular phone-related topics. However, because there is no information about the entity ``Galaxy Nexus" in the ODP, the ODP-based classifier would misclassify the text into the ODP category \textit{/Science/../Galaxies}. 

In this paper, we propose a new approach to enrich the semantics of ODP categories with entities, which facilitates the effectiveness of the ODP-based classification. To enrich ODP categories with entities, we leverage a knowledge base called Probase\footnote{https://concept.research.microsoft.com}. Probase is a probabilistic knowledge base and it contains millions of entities and concepts. One of the advantages of Probase is that in comparison with the well-known knowledge bases, such as WordNet\footnote{\url{https://wordnet.princeton.edu}} or Yago\footnote{\url{http://www.yago-knowledge.org}}, it has high taxonomy coverage\footnote{The taxonomy coverage means that how much the taxonomy contains at least one term.} of texts, which enables it to understand the semantic meaning of the text more thoroughly \cite{wu2012probase}. To the best of our knowledge, this is the first work to combine the two well-known knowledge bases, ODP and Probase, for the ODP-based text classification.

\begin{figure*}[h] 
 \begin{center} 
 \includegraphics [width=0.85\textwidth]{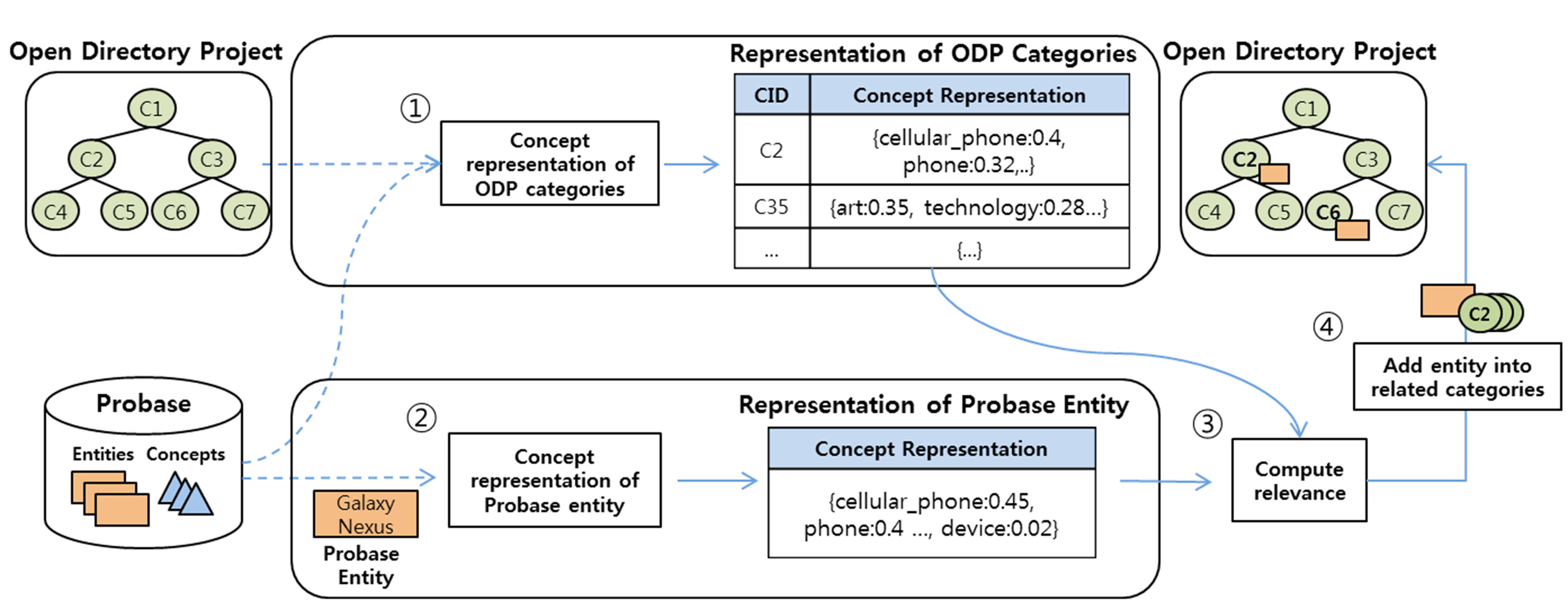}
 \caption{Overview of Our Methodology}
 \label{overview}
 \end{center}
\end{figure*}

Figure \ref{overview} provides an overview of our methodology. To enrich the semantics of ODP categories with Probase entities, we first represent ODP categories and Probase entities in terms of concepts (refer to \textcircled{\small{1}} and \textcircled{\small{2}} in Figure \ref{overview}). Next, we measure the semantic relevance between the ODP category and the Probase entity using concept representations (refer to \textcircled{\small{3}} in Figure \ref{overview}). We compute the semantic relevance for all categories and re-scale this scores as a probability. Finally, based on the probability, we add the entity into the top-$k$ related ODP categories (refer to \textcircled{\small{4}} in Figure \ref{overview}).

To illustrate our methodology, we assume that there is a Probase entity ``Galaxy Nexus". We represent ``Galaxy Nexus" with concepts, such as ``phone" or ``cellular\_phone". Similarly, we represent each ODP category with concepts, e.g., the ODP category \textit{/Shopping/.../Cellular\_Phone} with concepts, such as ``phone" or ``cellular\_phone". Next, we measure the semantic relevance between the ``Galaxy Nexus" and each ODP category, utilizing concept representations. Finally, we add the ``Galaxy Nexus" into the top-$k$ related ODP categories, e.g., the phone-related ODP categories, such as \textit{/Shopping/.../Cellular\_Phone}, based on the relevance score.
 
The technical contributions of our work are summarized as follows:
\begin{itemize}
\item We improve the ODP-based text classification by enriching the semantics of ODP categories with Probase entities. 
\item We propose a new approach to represent each ODP category and Probase entity as a concept representation to measure the semantic relevance between an ODP category and a Probase entity.
\item We demonstrate the effectiveness of our methodology by evaluating the classification performance. The results clearly show that our methodology significantly outperforms current state-of-the-art techniques.
\end{itemize}

The remainder of this paper is structured as follows. In Section \ref{sec:preliminary}, we introduce knowledge bases and the classification framework adopted from the work \cite{lee2013semantic}, \cite{ha2014toward_MCAD}. We then describe our proposed methodology in Sections \ref{sec:shared_info}, \ref{sec:con_representation} and \ref{sec:relevant}. We show and analyze the performance results in Section \ref{sec:experiment}. We discuss related work in Section \ref{sec:related} and conclude in Section \ref{sec:conclusion}.

\section{Preliminary\label{sec:preliminary}}
\subsection{Probase}

\begin{figure}[h]
	\begin{center} 
		\includegraphics[width=0.35\textwidth]{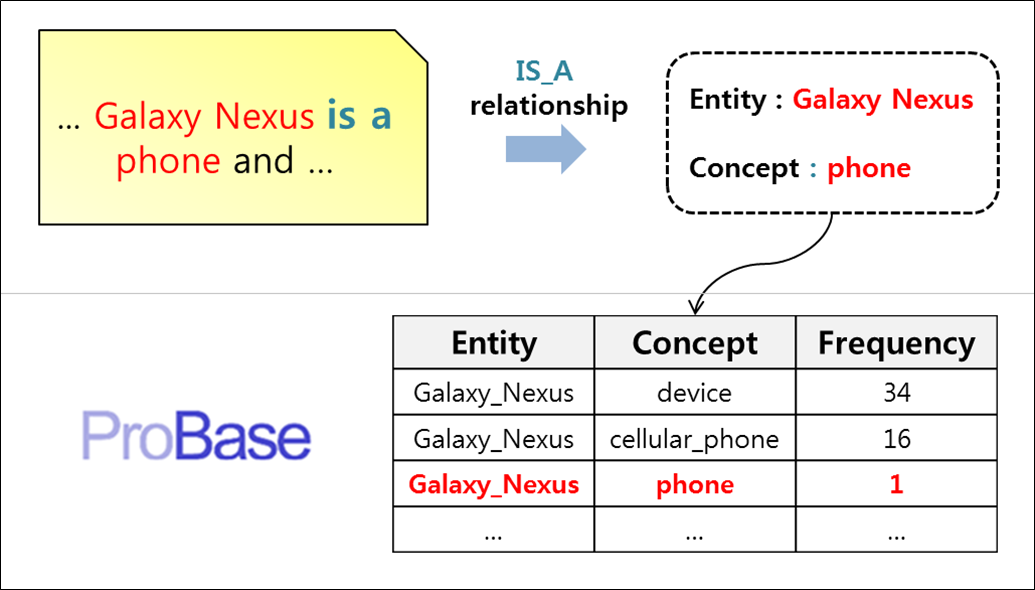}%
		\caption{(a) Structure of Probase 
		\label{probase_ex}} 
		\par\medskip 
	\end{center}
\end{figure}

Probase in Figure \ref{probase_ex} is an automatically constructed knowledge base, built with linguistic patterns derived from billions of web pages. Probase contains millions of entities and concepts and formalized relationships between entities and their enclosing concept classes with their co-occurrence frequency, e.g., entity ``Galaxy Nexus" and concept ``smartphone". Probase data are publicly available; thus, many researchers have utilized this knowledge base for tasks including text conceptualization \cite{song2011short}, \cite{wang2015query}, short text classification \cite{similar_mine} and taxonomy construction \cite{song2015automatic}. 
 
\subsection{Open Directory Project}

\begin{figure}[h]
	\begin{center} 
		\includegraphics[width=0.35\textwidth]{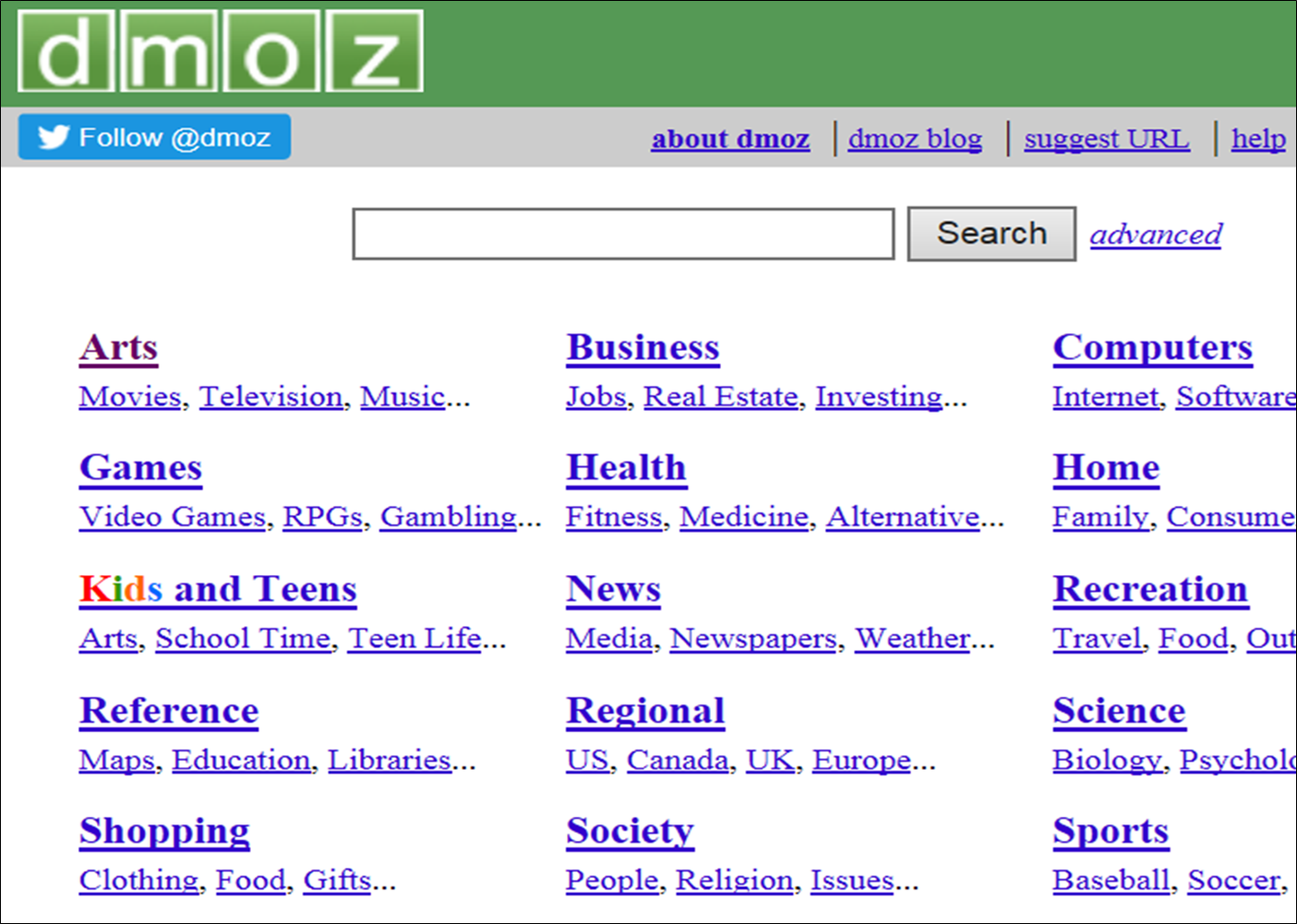}%
		\caption{(b) Structure of ODP 
		\label{fig_kb}} 
		\par\medskip 
	\end{center}
\end{figure}

ODP in Figure \ref{fig_kb} is a large-scale web directory built by human editors. It contains approximately one million ODP categories and millions of web pages pre-classified into categories. 
The ODP categories have a hierarchical structure; there is an \textit{IS-A} relationship between a parent and a child category. ODP data are publicly available; thus, it is widely applied to various research areas, such as web classification \cite{shin2017utilizing}, \cite{ha2014toward_MCAD}, contextual advertising \cite{lee2013semantic}, \cite{ryu2017utilizing}, and personalized curation \cite{ODP_Mobile1}.

\subsection{ODP-based Classification} 
We employ \cite{lee2013semantic} for constructing the ODP-based text classifier. The ODP is a large-scale web directory built by human editors. The ODP categories have a hierarchical structure, which has \textit{IS-A} relationship between a parent and a child category \cite{lee2013semantic,isa:98}. 
To train the ODP-based text classifier, we compute the centroid $\overrightarrow{\mu (t) }$ of ODP category $t$ by averaging all term vectors across all ODP documents as follows:
\begin{equation}
\label{centroid}
\begin{split}
\overrightarrow{\mu (t) } = \frac{1}{\left| D_{t} \right|}\sum_{d\in D_{t}}^{}\overrightarrow{v(d) }, 
\end{split}
\end{equation}
where $D_{t}$ is a set of ODP documents in ODP category $t$ and $\overrightarrow{v(d)}$ is a weighted vector represented as a \textit{tf-idf} value. However, some ODP categories have the data sparsity problem, since 58\% of categories have less than five texts that classified into themselves \cite{lee2013semantic}. Therefore, we employ \cite{lee2013semantic,ha2014toward_MCAD} that has merged the centroid vector $\overrightarrow\mu (t) $ of the ancestor and descendant ODP categories to build a classifier. As a result, it alleviate the data sparsity problem in the ODP.

\section{Representing Shared semantics of ODP category and Probase entity \label{sec:shared_info}}

In this paper, we enrich the semantics of ODP categories with Probase entities, which facilitates the effectiveness of the ODP-based classification. To do so, we add semantically related Probase entities to ODP categories by measuring the semantic relevance between them. Thus, it is a challenge to measure the relevance between ODP category and Probase entity, since there is a structural difference in the heterogeneous knowledge bases. The ODP consists of categories and documents, whereas Probase consists of concepts and entities. Therefore, it is demanding to find a common representation of semantic information that can be shared between them.

As a solution to this problem, we represent ODP categories and Probase entities in terms of concepts. In this paper, we define the concept as a class of entities or categories within the domain \cite{similar_mine}. For example, the Probase entity ``Galaxy Nexus" can include a set of concepts such as ``phone" or ``cellular\_phone". Likewise, the ODP category \textit{/Shopping/../Cellular\_Phone} can include the concepts such as ``phone" or ``cellular\_phone". Thus, if an ODP category and a Probase entity belong to similar classes, their concept representations will be similar. Based on this observation, we represent ODP categories and Probase entities in terms of concept vectors and measure the semantic relevance between them.

\section{Concept Representation of ODP Categories} \label{sec:con_representation}

In this section, we describe how to represent ODP categories in concepts (refer to \textcircled{\small{1}} in Figure \ref{overview}). Specifically, we introduce a way for searching for concepts in ODP categories and representing them as concept vectors. We then explain how to enrich the diversity of concepts in each ODP category to better represent its semantics. 

\subsection{Representing ODP Category with Concept Vector}
We represent ODP categories using concepts. However, there is no concept information in the ODP, which makes difficult to represent concept vectors of ODP categories. Thus, we propose a methodology for finding related concepts in each ODP category. Given a set of ODP documents, we break down the ODP document into text segments. Then, we match syntactic relationships between segments and concepts in Probase. We regard these segments as candidate concepts in the ODP category. 

After searching for concepts in ODP documents, we measure representativeness of each candidate concept in ODP category. To measure representativeness, we apply the $tf$-$idf$ scheme. Specially, we use $tf$ because concepts of high frequency in ODP document are likely to be more important and descriptive for the ODP category. Likewise, we use $df$ because it discriminates the degree of semantic importance of a concept. We use $df$ from the documents in both of the knowledge bases, the ODP and Probase. We compute a concept weight $w(c_{t})$ in the ODP category $t$ as follows:
\begin{equation}
\label{weighted_category_vector}
\begin{split}
w\left ( c_{t} \right) = cw_{odp} \cdot log\left ( cw_{odp} \right) \cdot cw_{pro} \cdot log\left ( cw_{pro}\right), 
\end{split}
\end{equation}
where 
\begin{equation}
\label{weighted_category_and_entity}
\begin{split}
cw_{odp} = \frac{tf \left ( c_{t} \right) }{log_{}\left ( df_{odp}\left ( c_{t} \right) \right) }, cw_{pro} = \frac{tf \left ( c_{t} \right) }{log_{}\left ( df_{pro}\left ( c_{t} \right) \right) }.
\end{split}
\end{equation}
$cw_{odp}$ and $cw_{pro}$ are the $tf$-$idf$ based concept weights in category $t$. $df_{odp}$ and $df_{pro}$ is the number of source documents from ODP and Probase where concept $c_{t}$ appears. $tf \left (c_{t}\right)$ is a term frequency of concept $c_{t}$ appeared in ODP category $t$. We take the logarithm to make concepts with low frequencies have low weights in the ODP concept vector. 

\subsection{Enriching Concept Information of ODP Category}
When we represent ODP categories as concept vectors, we encounter the scarcity problem of concept lists in each ODP category. This problem is caused by the deficiency of training data in each ODP category. In our experiment regarding the ODP, approximately 72\% of ODP categories have less than five documents classified into themselves. The deficiency of training documents leads to the poor concept representation of an ODP category, which complicates the measurement of the relatedness between an ODP category and a Probase entity.

As a solution to this problem, motivated by \cite{lee2013semantic}, \cite{ha2014toward_MCAD}, we increase the number of concepts in an ODP category using the hierarchy structure of the ODP. 
Specifically, we increase the number of concepts that are found in the descendants of a specific ODP category. Thus, we enrich the diversity of a concept vector, $\overrightarrow{c(t)^\prime}$ in the ODP category $t$ as follows:

Let $\overrightarrow{c(t)^\prime}$ be the enriched concept vector of ODP category $t$, we merge the concept vector of the descendant ODP categories as follows:
\begin{equation}
\label{merge_concept}
\begin{split} 
\overrightarrow{c (t) ^{\prime}} = \alpha\cdot\overrightarrow{c (t) } + (1-\alpha) \cdot\frac{1}{\left| child (t) \right|}\sum_{t_{k}\in child(t)}^{}\overrightarrow{c (t_{k}) ^{\prime}},
\end{split}
\end{equation}
where $child(t)$ is a set of child categories of the ODP category $t$ and $\overrightarrow{c(t)}$ is a concept vector of the ODP category $t$. Enriched concept vector $\overrightarrow{c(t)^\prime}$ is a linear combination of concept vector $\overrightarrow{c(t)}$ and the sum of enriched concept vectors of child categories of ODP category $t$.

\section{Enriching ODP Categories with Probase Entities\label{sec:relevant}}
In this section, we describe how to enrich the semantics of ODP categories with Probase entities. Specifically, we introduce how to represent Probase entities with concepts. We then measure the semantic relevance between an ODP category and a Probase entity. Finally, we add Probase entities to the related ODP categories based on the semantic relevance.

\subsection{Concept Representation of Probase Entity}

The second part of the methodology (refer to \textcircled{\small{2}} in Figure \ref{overview}) is to represent Probase entities with concepts. Given a specific Probase entity, we obtain a list of related concepts, which are already provided in Probase. For example, the related concepts of the Probase entity ``Galaxy\ Nexus" include ``smartphone", ``product" or ``multi\_touch\_phone". Among these concepts, however, humans are less likely to associate the entity ``Galaxy\ Nexus" with general concepts, such as ``product" because they are used with many entities. Likewise, specific concepts, such as ``multi\_touch\_phone" cannot be regarded as representative because they are not frequently used with entities. 

To assign appropriate concepts to entities, many researchers \cite{wang2015query}, \cite{lee2013attribute}, \cite{wang2015inference} rely on a probabilistic approach called \textit{typicality}. In this paper, we borrow the \textit{typicality} proposed by \cite{wang2015inference} as a score function. This function assigns a high representative score to concepts that are not too general nor specific. We compute typicality score $w\left (c_{e}\right)$ of the Probase entity $e$ as follows:
\begin{equation}
\label{typicality}
\begin{split}
w\left ( c_{e} \right) = P\left ( e|c \right) \cdot P\left ( c|e \right), 
\end{split}
\end{equation}
where 
\begin{equation}
\label{typicality}
\begin{split}
P\left ( e|c \right) = \frac{n\left ( c, e \right) }{\sum_{e_{i}\in c}^{} n\left ( c, e_{i} \right) }, 
\qquad
P\left ( c|e \right) = \frac{n\left ( c, e \right) }{\sum_{e\in c_{i}}^{} n\left ( c_{i}, e \right) }.
\end{split}
\end{equation}
$n\left(c, e \right)$ is the co-occurrence of $c$ and $e$. This frequency information is already provided in Probase. $P(e|c)$ is the typical score of Probase entity $e$ in Probase concept $c$, and it assigns higher scores to specific concepts of the entity $e$. In contrast, $P(c|e)$ is the typical score of Probase concept $c$ for entity $e$, and it assigns higher scores to more general concepts of entity $e$. Thus, consideration of both $P(e|c)$ and $P(c|e)$ facilitates assigning a high representative score to concepts that are not too general or specific. After scoring the typicality of all concepts of the entity $e$, we selectively choose the concepts, whose typicality scores exceed a certain threshold $\beta$, based on our preliminary experiments. 

\subsection{Relevance between ODP Category and Probase Entity}

The last part of the methodology (refer to \textcircled{3} in Figure \ref{overview}) is to measure the semantic relevance between ODP category and Probase entity. We compare concept vectors of an ODP category and a Probase entity. For example, the ODP category \textit{/Shopping/../Cellular\_Phone} and the Probase entity ``Galaxy\ Nexus" are similar, since they share common concepts, such as ``phone" or ``cellular\_phone". 

Given an ODP category $t$ and a Probase entity $e$, we define the semantic relevance score $rel(t,e)$ between ODP category and Probase entity as follows:
\begin{equation}
\label{relevance_score}
\begin{split}
rel\left ( t, e \right) = \sum_{c_{t}\in T}^{} \sum_{c_{e}\in E }^{}sim\left (c_{t}, c_{e} \right) \cdot w\left ( c_{t} \right) \cdot w\left ( c_{e} \right),
\end{split}
\end{equation}
where $T$ is a set of concepts in concept vector of the ODP category $t$, and $E$ is a set of concepts in concept vector of Probase entity $e$. In addition, $sim (c_{t}, c_{e})$ is a similarity score between two concepts $c_{t}$ and $c_{e}$, which is already provided in Probase. $w(c_{t})$ is concept vector of ODP category $t$ and $w(c_{e})$ is concept vector of Probase entity $e$.

Next, we measure the semantic relevance for all categories and re-scale this scores as a probability using softmax. Based on this probability, we rank ODP categories for the Probase entity and select the top-$k$ ODP categories as related categories for the Probase entity. Finally, we add the Probase entity into the related ODP categories (refer to \textcircled{\small{4}} in Figure \ref{overview}). 

\section{Performance Experiment\label{sec:experiment}}
 \subsection{Dataset}
Table I shows statistics of the datasets and Table \ref{param-table} shows the selected parameter values. These values are determined empirically. 

\subsubsection{ODP Dataset}
We use the ODP RDF dump, released on October 2014, from the original ODP dataset. It contains 796,902 ODP categories and 3,917,043 web pages. To obtain a well-organized taxonomy, we apply heuristic rules described in \cite{lee2013semantic}. As a result, we use 4,521 categories to build a taxonomy.
 
\begin{table}[!t]
\renewcommand{\arraystretch}{1.1}
\caption{Statistics of Datasets}
\centering
\begin{tabular}{|c||c|c|c|}\hline  
\multirow{4}{*}{Training Dataset} &\multirow{2}{*}{ODP} & Categories & 796,902 \\ \cline{3-4}
& & Documents & 3,917,043 \\ \cline{2-4}
& \multirow{2}{*}{Probase} & Entities & 6,215,858 \\ \cline{3-4}
& & Concepts & 2,359,856 \\ \hline
\multirow{2}{*}{Test Dataset} & Probase Entity & Entities & 115 \\ \cline{2-4}
 & News & Texts & 100 \\ \hline
\end{tabular}
\label{dataset}
\end{table} 

\subsubsection{Probase Dataset}
We use the Probase dataset, released on July 2013, to enrich the ODP-based text classification. It contains 6,215,858 entities and 2,359,856 concepts. To utilize only representative concepts for Probase entities, we choose the concepts whose the typicality scores ($\beta$) exceed 0.004, based on our preliminary experiments. Then, we select Probase entity that has at least one concept representing the Probase entity. As a result, 1,000,500 Probase entities are used for enriching the semantics of ODP categories.

\subsubsection{Probase Entity Dataset}
We use Probase entity dataset to evaluate the matching performance between ODP categories and entities. We evaluate this to show that our proposed methods helps to enrich the ODP categories with the semantically related entities. This dataset is randomly selected 115 entities from Probase, covering different topics such as fashion, movie, sports, and health. 

\subsubsection{News Dataset}
We use New York Times (NYT) news dataset to evaluating the classification performance on a real-world dataset. We select five categories: art, business, fashion, movie, and sports in the news categories. Then, we randomly collect approximately 20 news articles from these news categories, where each news article includes at least one Probase entity. We use precision at $k$ as an evaluation metric in the same fashion as \cite{shin2017utilizing}.
\begin{table}[h]
\renewcommand{\arraystretch}{1.1}
\caption{\label{param-table} Parameter Setting}
\centering
\begin{tabular}{|c|c|c|}
\hline Notation & Meaning & Value \\ 
\hhline{|=|=|=|} 
$\alpha$ & merge ratio & 0.7 \\ \hline 
$\beta$ & typicality threshold & 0.004 \\ \hline
$\tau$ & weight of entity & 0.8 \\ \hline 
$k$ & top-$k$ categories & 5 \\ \hline
\end{tabular}
\end{table} 

\subsection{Experimental Setup}
For the assessment, three researchers manually assess the ODP categories obtained by the ODP-based text classifiers according to three scales: relevant, somewhat relevant and not relevant. We use precision at $k$ as an evaluation metric for both of the Probase Entity dataset and the news dataset. For each entity or news, we manually annotate the top-$k$ categories selected by each method and we measure the precision at each position $k$. For example, if three categories out of the top-five categories are relevant, then the precision at five is measured to be 0.6.

We evaluate the performance of the following five models:
\begin{itemize}
\item \textit{ODP}: The ODP-based text classifier \cite{ha2014toward_MCAD}. It is the baseline for the ODP-based text classification.
\item \textit{ODP + Wiki}: The ODP-based text classification enriched with Wikipedia phrases and relevant hyperlinks \cite{shin2017utilizing}. It is the state-of-the-art method for the ODP-based text classification.
\item \textit{ODP + $Probase_{path}$}: The ODP-based text classification enriched with Probase entities. This method obtains concepts of each ODP category from its path information. It represents ODP categories as the $tf$ based concept vectors.
\item \textit{ODP + $Probase_{tf}$}: The ODP-based text classification enriched with Probase entities. This method obtains concepts of each ODP category from web pages in the ODP category. It represents ODP categories as the $tf$ based concept vectors.
\item \textit{ODP + $Probase_{tf\_idf}$} (proposed model): The ODP-based text classification enriched with Probase entities. This method obtains concepts of each ODP category from web pages in the ODP category. It represents ODP categories as the $tf$-$idf$ based concept vectors.
\end{itemize}

\subsection{Experimental Results} 
\subsubsection{Parameter Setting} 

Table \ref{param-table} shows the parameters of our methodology. We set the merge ratio ($\alpha$) as 0.7 to generate merged-concept vectors. We selectively choose the concepts whose the typicality scores ($\beta$) exceed 0.004, based on our preliminary experiments. If typicality score of the concept is less than 0.004, it means the concept is too general or specific concept.
 
\begin{figure}[h]
 \begin{center} 
 \includegraphics [width=0.45\textwidth]{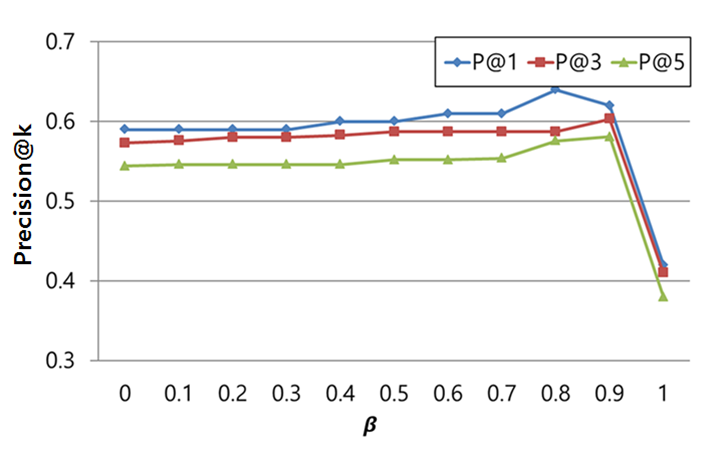}
 \caption{Classification Performance on News Dataset (4,521 ODP categories) based on Different $\tau$ (weight of entity word)}
 \label{para_combining}
 \end{center}
\end{figure}

We use the parameter $\tau$ to determine the optimal ratio of Probase entities and ODP words when we classify the text. We set parameters as follows: $\tau$ $\times$ Probase entity + (1-$\tau$) $\times$ ODP words. Figure \ref{para_combining} shows the classification performance based on different parameters. We find that the performance of text classification increases as $\tau$ increased, up to 0.8; the curve reaches a peak at $\tau$ = 0.8. 
We observe that raising the importance of entity improves text classification performance. In addition, we find that when $\tau$ = 1.0, the performance significantly decreased. This suggests that both ODP terms and added Probase entities contribute largely to the ODP-based text classification. In the rest of the experiments, we therefore set $\tau$ to 0.8. 

\begin{table} [h]
\renewcommand{\arraystretch}{1.1}
\centering
\caption{Matching Results on Probase Entity Dataset (4,521 ODP categories)}
\label{performance_pro} 
\begin{tabular}{|c|c|c|c|}
\hline 
Method & Precision@1 & Precision@3 & Precision@5 \\ \hhline{|=|=|=|=|} 
\textit{ODP + Wiki} \cite{shin2017utilizing} & 0.505 & 0.460 & 0.455 $\ddagger$\\ \hline
\textit{ODP + $Probase_{path}$} & 0.285 $\ddagger$ & 0.339 $\ddagger$ & 0.322 $\ddagger$ \\ \hline
\textit{ODP + $Probase_{tf}$} & 0.171 $\ddagger$ & 0.178 $\ddagger$ & 0.164 $\ddagger$ \\ \hline
\textit{ODP + $Probase_{tf\_idf}$} & \textbf{0.590} & \textbf{0.578} & \textbf{0.568} \\ \hline
\end{tabular}
\end{table}  

\subsubsection{Results on Matching Performance} 
Table \ref{performance_pro} shows the matching performance on Probase entity dataset. In this experiment, we exclude the matched result of \textit{ODP} method because this method only utilizes the ODP knowledge base without adding entities. In result, our proposed method \textit{ODP+$Probase_{tf\_idf}$} outperforms \textit{ODP+Wiki} by 17\%, 26\%, and 25\% on precision at one, precision at three, and precision at five, respectively. This implies that our proposed method helps to enrich the ODP categories with the semantically related entities, since it grasps the semantic relationship between ODP category and Probase entity very well. 

In addition, compared with \textit{ODP + $Probase_{path}$}, \textit{ODP + $Probase_{tf\_idf}$} achieves much better performance, which implies that increasing the diversity of concepts in each ODP category helps to represent ODP categories more semantically. 
Lastly, compared with \textit{ODP + $Probase_{tf}$}, \textit{ODP + $Probase_{tf\_idf}$} achieves much better performance. It demonstrates the effectiveness of adopting document frequency when we represent ODP categories as a concept vectors.

To demonstrate statistical significance, we also perform the $t$-test on the classification results. The dagger symbol ($\dagger$) indicates a $p$-value$<$0.05, while the double dagger symbol ($\ddagger$) indicates a $p$-value$<$0.01. 

\begin{table} [h]
\renewcommand{\arraystretch}{1.1}
\centering
\caption{Classification Results on News Dataset (4,521 ODP categories)}
\label{performance_nyt} 
 \begin{tabular}{|c|c|c|c|} \hline
    {Method}    & Precision@1 & Precision@3 & Precision@5 \\ \hhline{|=|=|=|=|} 
    \textit{ODP} \cite{ha2014toward_MCAD}  & 0.550 $\dagger$ & 0.533 & 0.494 \\ \hline
    \textit{ODP + Wiki}  \cite{shin2017utilizing} & 0.590 & 0.566 & 0.572 $\dagger$ \\ \hline
    \textit{ODP + $Probase_{tf\_idf}$} & \textbf{0.640} & \textbf{0.583} & \textbf{0.576} \\ \hline
  \end{tabular}
\end{table}   

\subsubsection{Results on News Dataset} 
Table \ref{performance_nyt} shows the classification performance on news dataset. In result, the proposed method \textit{ODP + $Probase_{tf\_idf}$} outperforms all the other methods. It shows that enriching the semantics of ODP categories with Probase entities facilitates the large-scale text classification.
More specifically, \textit{ODP + $Probase_{tf\_idf}$} and \textit{ODP + Wiki} have better performance than \textit{ODP}, which means additional knowledge is useful to improve the performance of the ODP-based text classification. Morever, the proposed method \textit{ODP + $Probase_{tf\_idf}$} has better performance than \textit{ODP + Wiki}. It means that enriching the ODP categories with the semantically related entities improves actual classification performance. We give a more detailed qualitative analysis in the next subsection.
  
\subsection{Qualitative Analysis}
Table \ref{q_entity} shows the matching results for entity ``Galaxy Nexus". The bold ODP category is the correctly matched ODP category with the entity. In matching result from \textit{ODP + Wiki}, there are two correctly matched ODP categories for ``Galaxy Nexus". In contrast, all ODP categories from \textit{ODP + $Probase_{tf\_idf}$} are correctly matched ODP categories with the entity. It shows that \textit{ODP + $Probase_{tf\_idf}$} grasps the semantic relationship between ODP category and Probase entity better than \textit{ODP + Wiki}.
 
Table \ref{q_query} shows the classification result for the query ``Galaxy Nexus Spec" including the entity ``Galaxy Nexus". In the \textit{ODP} result, it misclassifies the text into science-related ODP category because there is no information about the entity ``Galaxy Nexus" in ODP. In contrast, \textit{ODP + $Probase_{tf\_idf}$} or \textit{ODP + Wiki} correctly classify the text into phone-related categories since they can understand the semantic meaning of ``Galaxy Nexus". Thus, adding entities to the relevant ODP category facilitates the text classification. In th e case of the proposed method \textit{ODP + $Probase_{tf\_idf}$}, there are more correctly classified categories compared with \textit{ODP + Wiki}. It shows that our proposed model grasps the semantic relationship between ODP category and Probase entity very well and it improves text classification performance. 
\begin{table*}[h]
\renewcommand{\arraystretch}{1.1}
\centering
\caption{Illustration of Matched Top-5 ODP Categories of Entity ``Galaxy Nexus"}
\label{q_entity} 
  \begin{tabular}{|c||c|c|l|}  \hline
  Entity & Method & Rank & \multicolumn{1}{c|}{Category} \\ \hline
  \multirow{10}{*}{Galaxy Nexus} &   \multirow{5}{*}{\textit{ODP + Wiki} \cite{shin2017utilizing}} & 1 & \textbf{\textit{/Shopping/Consumer\_Electronics/Communications/Wireless/Cellular\_Phones}}\\ \cline{3-4}
 & & 2 & \textbf{\textit{/Shopping/Consumer\_Electronics/Communications/Wireless}}\\ \cline{3-4}
 & & 3 & \textit{/Computers/Internet/Searching/Search\_Engines/Google}\\ \cline{3-4}
 & & 4 & \textit{/Computers/Software/Operating\_Systems}\\ \cline{3-4}
 & & 5 & \textit{/Computers/Internet/Searching/Search\_Engines}\\ \cline{2-4}
 
 &   \multirow{5}{*}{\textit{ODP + $Probase_{tf\_idf}$}} & 1 & \textbf{\textit{/Shopping/Consumer\_Electronics/Communications/Wireless}}\\ \cline{3-4}
 & & 2 & \textbf{\textit{/Business/Consumer\_Goods\_and\_Services/Electronics}}\\ \cline{3-4}
 & & 3 & \textbf{\textit{/Shopping/Consumer\_Electronics/Communications}}\\ \cline{3-4}
 & & 4 & \textbf{\textit{/Shopping/Consumer\_Electronics/Accessories}}\\ \cline{3-4}
 & & 5 & \textbf{\textit{/Shopping/Consumer\_Electronics/Communications/Wireless/Cellular\_Phones}}\\ \hline
  \end{tabular}
\end{table*}

\begin{table*}[h]
\renewcommand{\arraystretch}{1.1}
\centering
\caption{Illustration of Classified ODP Categories of Text ``Galaxy Nexus Spec"}
\label{q_query} 
  \begin{tabular}{|c||c|c|l|}  \hline
  Input Text & Method & Rank & \multicolumn{1}{c|}{Category} \\ \hline
  \multirow{15}{*}{Galaxy Nexus Spec} & \multirow{5}{*}{\textit{ODP} \cite{ha2014toward_MCAD}} & 1 & \textit{/Sports/Motorsports/Auto\_Racing/Organizations/SCCA}\\ \cline{3-4}
  & & 2 & \textit{/Games/Video\_Games/Action/Space\_Combat/Massive\_Multiplayer\_Online}\\ \cline{3-4}
  & & 3 & \textit{/Computers/Performance\_and\_Capacity}\\ \cline{3-4}
  & & 4 & \textit{/Games/Video\_Games/Action/Space\_Combat/Star\_Wars\_Games/Star\_Wars\_Galaxies}\\ \cline{3-4}
  & & 5 & \textit{/Science/Astronomy}\\ \cline{2-4}
  & \multirow{5}{*}{\textit{ODP + Wiki} \cite{shin2017utilizing}} & 1 & \textit{/Computers/Internet/Searching/Search\_Engines}\\ \cline{3-4}
  & & 2 & \textbf{\textit{/Shopping/Consumer\_Electronics/Communications/Wireless}}\\ \cline{3-4}
  & & 3 & \textit{/Computers/Internet/Searching/Search\_Engines/Google}\\ \cline{3-4}
  & & 4 & \textit{/Sports/Motorsports/Auto\_Racing/Organizations/SCCA}\\ \cline{3-4}
  & & 5 & \textit{/Science/Astronomy/Extraterrestrial\_Life}\\ \cline{2-4}
  & \multirow{5}{*}{\textit{ODP + $Probase_{tf\_idf}$}} & 1 & \textbf{\textit{/Shopping/Consumer\_Electronics/Communications/Wireless}}\\ \cline{3-4}
  & & 2 & \textbf{\textit{/Shopping/Consumer\_Electronics/Communications/Wireless/Cellular\_Phones}}\\ \cline{3-4}
  & & 3 & \textit{/Sports/Motosports/Auto\_Racing/Organizations/SCCA}\\ \cline{3-4}
  & & 4 & \textit{/Games/Video\_Games/Action/Space\_Combat/Star\_Wars\_Games/Star\_Wars\_Galaxies}\\ \cline{3-4}
  & & 5 & \textit{/Computers/Performance\_and\_Capacity}\\ \hline
  \end{tabular}
  \end{table*} 
  
\section{Related Work\label{sec:related}}
In text classification, many studies use machine learning techniques, such as support vector machine (SVM) \cite{joachims1998text_SVM} and Naive Bayes \cite{friedman1997bayesian} with bag-of-words (BoW) features. Due to the limitations of the BoW approach, many approaches have been developed for enriching semantic information by leveraging search engines or external knowledge base.


To obtain semantics information through search engines, \cite{shen2006query_searchengine} and \cite{sun2012short_searchengine} have expanded input text using search engines for classification. This approach differs from our proposed method in that it enriches the semantics of the input text. However, this expansion may not be suitable for real-time applications because it is very time consuming and heavily dependent on the quality of search engine \cite{chen2011short_searchengine_disa}.

In another line of work, several studies \cite{shin2017utilizing,rodriguez2000using_wordnet_classification} have enriched the semantics of categories in classifier using knowledge base, such as Wikipedia or WordNet. In particular, \cite{shin2017utilizing} has employed Wikipedia to enrich the semantics of ODP categories with phrases and hyperlinks for the ODP-based text classification. However, the coverage of additional semantic information is limited because the average number of Wikipedia phrases per ODP document is approximately 0.5 \cite{shin2017utilizing}. In addition, it might contain Wikipedia phrases, which have little relevance to the ODP category, such as Wikipedia phrase ``service plan" in the ODP category \textit{/Shopping/../Cellular\_Phone}. 
%
%

\section{Conclusion\label{sec:conclusion}}
In this paper, we have sought to enrich the semantics of ODP categories with Probase entities to better understand the semantics of text. Our proposed scheme has involved three tasks. First, we have represented each ODP category and Probase entity with concept representations. Second, we have measured the semantic relevance between an ODP category and a Probase entity. Finally, we have enriched ODP categories with related Probase entities based on the measured semantic relevance. We have verified the superiority of our proposed methodology in large-scale text classification on a real-world dataset. We plan to apply the proposed methodology to real-world applications, including contextual advertising and mobile advertising. 

\section*{Acknowledgement}
This research was supported by Basic Science Research Program through the National Research Foundation of Korea (NRF) funded by the Ministry of Science, ICT and Future Planning (number 2015R1A2A1A10052665).

\bibliographystyle{IEEEtran}
\bibliography{authors,IEEEabrv}

\end{document}